\documentclass[journal=jpcbfk,manuscript=article]{achemso}
\setkeys{acs}{usetitle = true}

\usepackage{color}

\author{Federico Coscio}
\affiliation[UCASAL, Salta, Argentina.]{Facultad de Arquitectura y Urbanismo, Universidad Cat\'olica de Salta, Salta, Argentina.}
\email{fcoscio@hotmail.com}
\phone{ (++54 +9387) 5351856}
\author{Alejandro D. Nadra}
\affiliation[UBA-CONICET-IQUIBICEN]{Instituto de Biociencias, Biotecnología y Biología Traslacional $IB^3$, Facultad de Ciencias Exactas y Naturales, Universidad de Buenos Aires-CONICET, Buenos Aires, Argentina..}
\author{Diego U. Ferreiro}
\email{ferreiro@qb.fcen.uba.ar}
\phone{ (++54 +11) 4576-3300 }
\affiliation[UBA-CONICET-IQUIBICEN]{Protein Physiology Lab, Dep de Qu\'imica Biol\'ogica, Facultad de Ciencias Exactas y Naturales, Universidad de Buenos Aires-CONICET-IQUIBICEN, Buenos Aires, Argentina.}

\title{A structural model for the Coronavirus Nucleocapsid}

\begin{document}
\begin{abstract}

We propose a mesoscale model structure for the coronavirus nucleocapsid, assembled from the high resolution structures of the basic building blocks of the N-protein, CryoEM imaging and mathematical constraints for an overall quasi-spherical particle. The structure is a truncated octahedron that accommodates two layers: an outer shell composed of triangular and quadrangular lattices of the N-terminal domain and an inner shell of equivalent lattices of coiled parallel helices of the C-terminal domain. The model is consistent with the dimensions expected for packaging large viral genomes and provides a rationale to interpret the apparent pleomorphic nature of coronaviruses.

\end{abstract}



\small

\section*{Introduction}
 {\it
``I want to find happiness in the tiniest of things and I want to try to do what I've been wanting to do for so long, that is, to copy these infinitesimally small things as precisely as possible and to be aware of their size.''} M.C. Escher

The coronavirus (CoV) N protein is a multifunctional element that binds the viral RNA and forms the major component of the ribonucleoprotein (RNP) forming the virion core. An important piece of missing information about CoVs lies in the difficulty in solving the atomic structure of the RNP complex, which has been hindered by its low solubility and the labile nature of the full-length N protein \cite{McBride:2014aa} . N protein is composed of three distinct and highly conserved domains: two structural and independently folded regions, N terminal domain (NTD, $\sim$180 amino acids) and C-terminal domain (CTD, $\sim$160 amino acids), that are separated by an intrinsically disordered region (LKR, $\sim$70 amino acids). Self-association of the N protein has been observed in many viruses, and is required for the formation of the viral capsid \cite{Cong:2017aa} . High resolution structures of both NTD and CTD in various crystal forms are available \cite{Berman:2000aa} . 

The current virion model of CoVs depicts a roughly spherical  pleomorphic particle that shows variations in size (80–120 nm) and shape (reviewed in \cite{Masters:2006aa}). The RNP is surrounded by a lipidic envelope of unusual thickness (7.8 nm), almost twice that of a typical biological membrane \cite{Barcena:2009aa} . Early studies showed that samples in which the viral envelope was disrupted and the inner content released,  RNPs from several CoVs appear to have helical symmetry \cite{Macneughton:1978aa} . However, under certain conditions, intermediate spherical nucleocapsids have been reported \cite{Garwes:1976aa, Risco:1996aa} .  These structures, which showed polygonal profiles, were devoid of S protein and lipids and, in addition to RNA and the N protein, contained the M protein. These assemblies were suggested to represent an additional viral structure: a shelled core, possibly even icosahedral \cite{Risco:1996aa} , that would enclose the helical RNP.

As was early noted by Crick and Watson, viruses should exhibit a high degree of symmetry as a consequence of ‘genetic economy’, the limited capacity in the viral genome to code for the proteins forming the capsid \cite{CRICK:1956aa} . Caspar and Klug extended this idea by introducing the principle of quasi-equivalence \cite{CASPAR:1962aa} , that allows larger viruses to form, requiring even smaller relative portions of their genomic sequences to code for their capsids. Recently, Twaroc and Luque showed that this theory is a special case of an overarching design principle for icosahedral, as well as octahedral, architectures that can be formulated in terms of the Archimedean lattices and their duals \cite{Twarock:2019aa} . The basic symmetry rules imply a general principle whenever a structure of a definitive size and shape has to be built up from smaller subunits \cite{Wolynes:1996aa} , the packing arrangements have to be repeated and hence the subunits are likely to be related by symmetry operations \cite{CRICK:1956aa} . Spherical viruses must be contained in one of the three possible cubic point groups: tetrahedron, cube/octahedron, dodecahedron/icosahedron. 

Here we built a mesoscale model for the CoVs nucleocapsid by bridging a bottom up approach from the high resolutions structures of NTD and CTD and a top-down approach from mathematical models of quasi-spherical particles.

\section*{Results}

\subsection*{Polyhedral approximations to the components}

Modularity is a necessary condition to construct form starting from basic building blocks \cite{Thompson1917} . Structural proteins must hence combine in modules that obey certain types of geometries and exist as monomers, dimers, trimers, etc \cite{Parra:2013aa} . There are limited possibilities to build structures that are close to spherical and that optimize resilience. These can be algebraically constructed and can be expressed as three-dimensional shapes with sets of regular polyhedra, its truncations, duals and geodesics. These polyhedra have three algebraic origins: the icosahedral, octahedral and tetrahedral, and its capacities to tessellate the bi and tri-dimensional space. As far as our current understanding goes, most of the viruses adopt an efficient strategy, that of the icosahedron, that can get close to sphericity with its 20 facets \cite{Rossmann:2013aa} . Examples of even more efficient forms include the icosidodecahedron (32 facets), and its dual, the rhombic triacontahedron (30 facets). The octahedral and the hexahedral geometries are algebraically bonded and are capable of generating a set of polyhedra by truncation and geodesic iteration that occupy the volume very close to a sphere. Given their algebraic origin, these are tillable by self- similar building blocks and can be modulated \cite{Coscio2014} . An overview of the model we propose is presented in Figure 1.

\medskip
	\begin{figure}
\centering
	\includegraphics[width=0.95\textwidth]{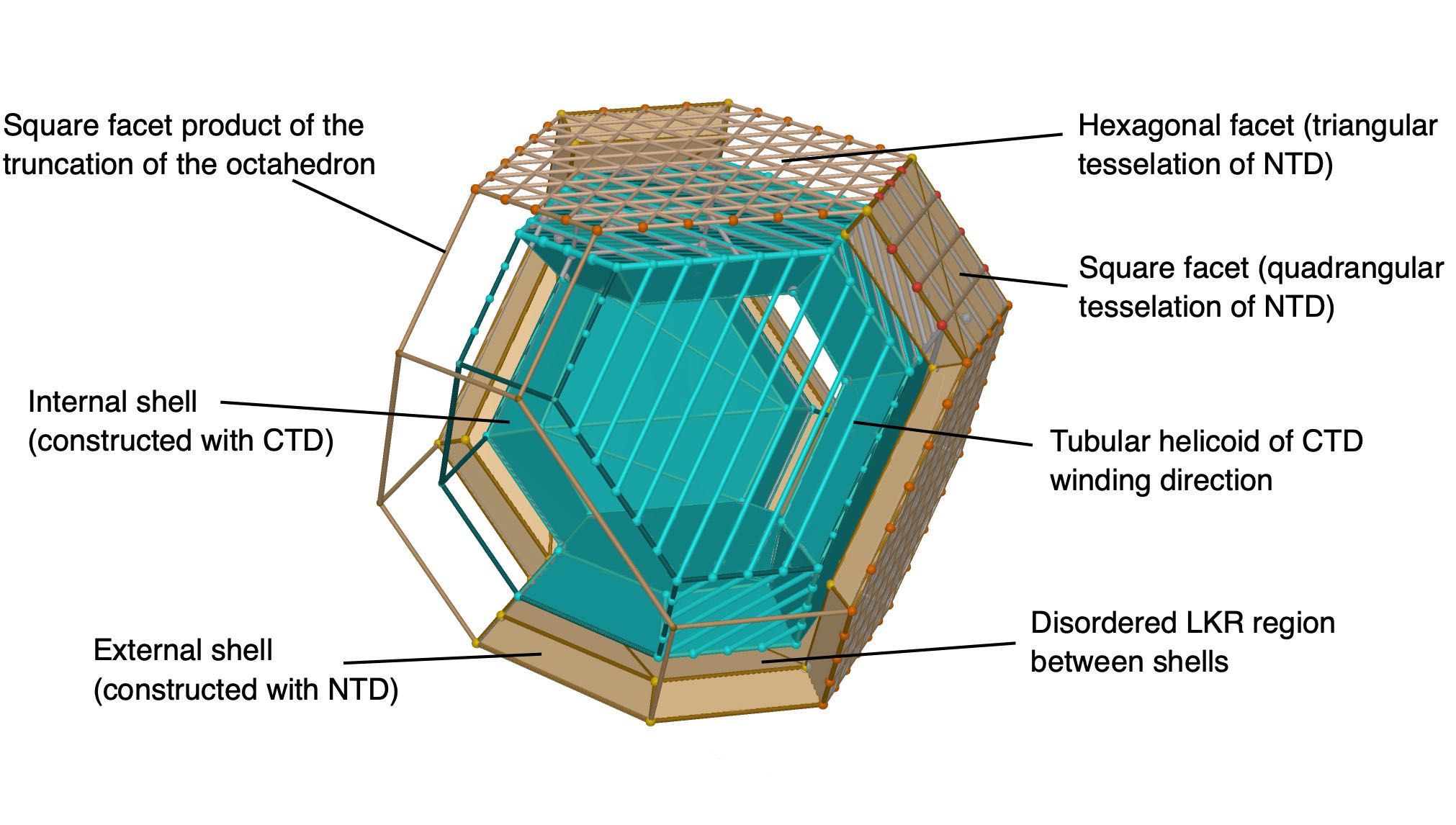}
	\caption{ Overall configuration of the nucleocapsid model. Two concentric shells of a truncated octahedron are built by the globular domains of the N protein: NTD tessellates the external facets and CTD packs the internal shell by winding a continuous coil. The disordered region LKR connects the two polyhedra.
	}
	\label{fig:crude}
\end{figure}
\medskip

There is a fundamental characteristic that gives the octahedral geometry a preference for CoVs RNP structure. This geometry allows the polyhedra to be constructed with a continuous linear coil winding, that is to say one linear element composed of identical modules that, in the case of RNP, is a large tubular helicoid that preserves the long, continuous, RNA chain. The truncated vertex of the octahedron is bidimensionally tillable in a linear and periodic manner. On the contrary, the pentagonal truncated vertex of the icosahedron is only tillable by non linear, aperiodic structures and thus cannot be built with a linear, modulated, coil winding. Accordingly, CTD can conform helices that have akin geometry to the octahedron, constructing threads that can turn at 135$^\circ$ preserving the helical constitution.

\subsection*{External shell}

The NTD (pdb 6M3M) has an asymmetric unit of $\sim$4.2nm along its longest axis, and can form a tetramer with an approximated triangular shape (Fig. 2b). Notably, NTD can also form a tetramer with a flat square shape (pdb 6VYO, Fig. 2c).

According to CryoEM experiments, the average diameter of the internal part of the mature CoVs virion is $\sim$100nm. Electron density was observed at a distance of $\sim$15nm from the membrane, with a depth of $\sim$25nm \cite{Neuman:2006aa} . For the external shell, the edge of the non-truncated octahedron is 70nm, a diagonal distance of 76nm, and a diameter of 98.9nm for the sphere that inscribes the octahedron. The ordering of the hexagonal facets of the truncated octahedron are modulated by triangles whose geometric centers are 5nm apart (green dots in Fig. 2). Each tetramer coincides with two triangles of the equilateral tessellation, that forms without superimposition in a complex form that leaves two monomers pointing outwards and two monomers inwards (Fig. 2b). The hexagonal facet is formed with 6x4 modules and the square facet by 4x4 modules (Fig. 1 and Fig. 2a), making a grand total of 2752 monomers of NTD. Triangular paracrystalline lattices were observed by CryoEM that are compatible with these dimensions \cite{Neuman:2006aa} .

\medskip
	\begin{figure}
\centering
	\includegraphics[width=0.8\textwidth]{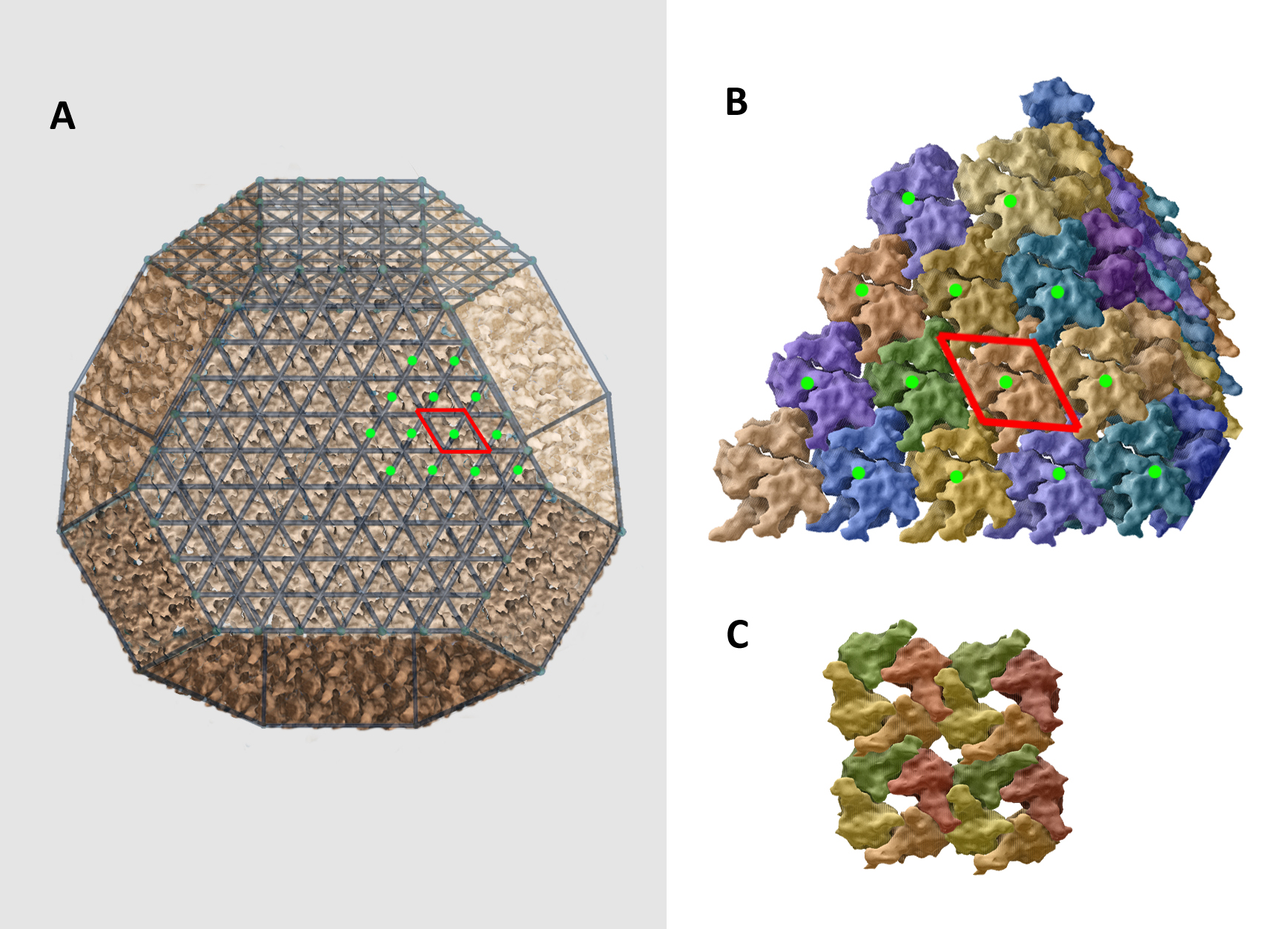}
	\caption{ Illustration of the external shell. A) NTD can tesselate both the hexagonal and the square facets of the truncated octahedron. B) high resolution structures of the NTD tetramer (pdb 6M3M) assembled in a triangular lattice (green dots). Two tetramers configure the rhomb marked in red. C) high resolution structure of NTD tetramer (pdb 6VYO) in the quadrangular configuration that tessellates the square facets.
	}
	\label{fig:crude}
\end{figure}
\medskip

\subsection*{Internal shell}

When disrupted, the virions release debris with tubular geometries of 9 to 15nm wide and hundreds of nm length \cite{Macneughton:1978aa, Barcena:2009aa} . Intact specimen observations show that the major part of the RNP locates $\sim$25nm on the internal side \cite{Neuman:2006aa} . Proteins near the viral membrane are arranged in overlapping lattices surrounding a disordered core \cite{Gui:2017aa}. Atomic densities at these depths revealed periodic patterns that can be interpreted as romboidal, triangular and quadrangular shapes, but no evidence of icosahedral geometry was found \cite{Neuman:2006aa, Gui:2017aa} . 

CTD domain is an obligated dimer, rotationally symmetric C1.2.1, with a major length of 5.6nm (pdb 2CJR). It can be configured in the form of an octamer of two anti parallel plates in a butterfly form of 10 nm wide, with an angle of $\sim$45$^\circ$ between dimers (Fig. 3a). This structure can also accommodate 2 types of rotations of 90$^\circ$ and 45$^\circ$ between modules (Fig. 3c). CTD can thus conform threads that have akin geometry to the octahedron, constructing helices that turn the direction in 135$^\circ$ preserving the helical constitution (Fig. 3c). The trihedral angle between the hexagonal and square facets of the truncated octahedron is precisely 135$^\circ$. An identical turn along the helix locates the tube  90$^\circ$ to the original direction (Fig. 3c).

The proposed structure for the packaging of RNP inside the virion consists of tubular arrangements of $\sim$10nm wide helices formed by CTD (Fig. 3d). These linear orderings are modulated in segments of $\sim$8nm for each CTD octamer and 4nm for the tetramers (Fig. 3a). The hexagonal facet is composed of 6x4 and the square facet by 4x4 modules (Fig. 1). There are 8 hexagons composed of 74 CTD tetramers and 6 squares composed of 16 CTD tetramers. The winding of the helicoid is completed with 2752 CTD, with a linear extension of RNP of $\sim$11000nm, coincident with the expected size for packing the complete CoV genome. The helicoidal assemblies that form the winding of the helices conform with the structure of the NTD outer shell in the form of a truncated octahedron, such that they contain n-1 triangular modules in each level of the hexagonal facets as they approach the vertices (Fig. 4). The packing is such that the tubular arrangements of 10nm stack 2.5nm, leaving 5nm for each level. 

\medskip
	\begin{figure}
\centering
	\includegraphics[width=0.8\textwidth]{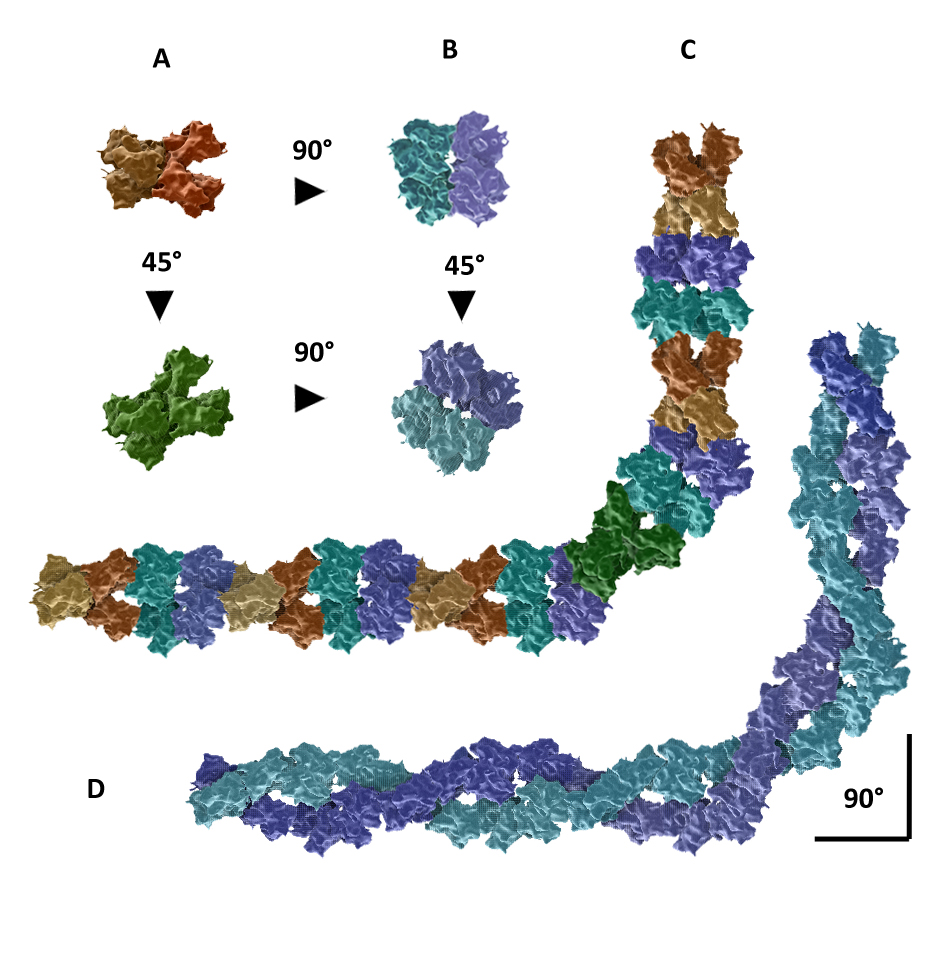}
	\caption{ Configuration of the internal shell. A) CTD octamers form a butterfly shaped module (pdb 2CJR) that is represented in different colors (B) according to their orientation. C) packing of CTD modules allow for the construction of helices that can turn 135$^\circ$ preserving the helical constitution. D) consecutive 135$^\circ$ turns locates the helix  90$^\circ$ to the original direction.
	}
	\label{fig:crude}
\end{figure}
\medskip

\subsection*{Relations in between shells}

Since the CTD modules are 4nm and the NTD modules are  5nm apart, the polyhedron constructed by NTD is 20\% larger. The thickness of the facets in the NTD shell is 4nm. Taking into account that the space in between the external facets of both polyhedra is 7nm, the interstitial space is about 3nm. NTD have 2nm protuberances to the inside that pack precisely against the tubular levels of CTD (Fig. 4). Since both NTD and CTD are connected by the LKR region, and both globular domains have been shown to interact with RNA \cite{McBride:2014aa} , we speculate that the interstitial space is filled with the LKR and RNA that is winded along the CTD helix.

\medskip
	\begin{figure}
\centering
	\includegraphics[width=0.95\textwidth]{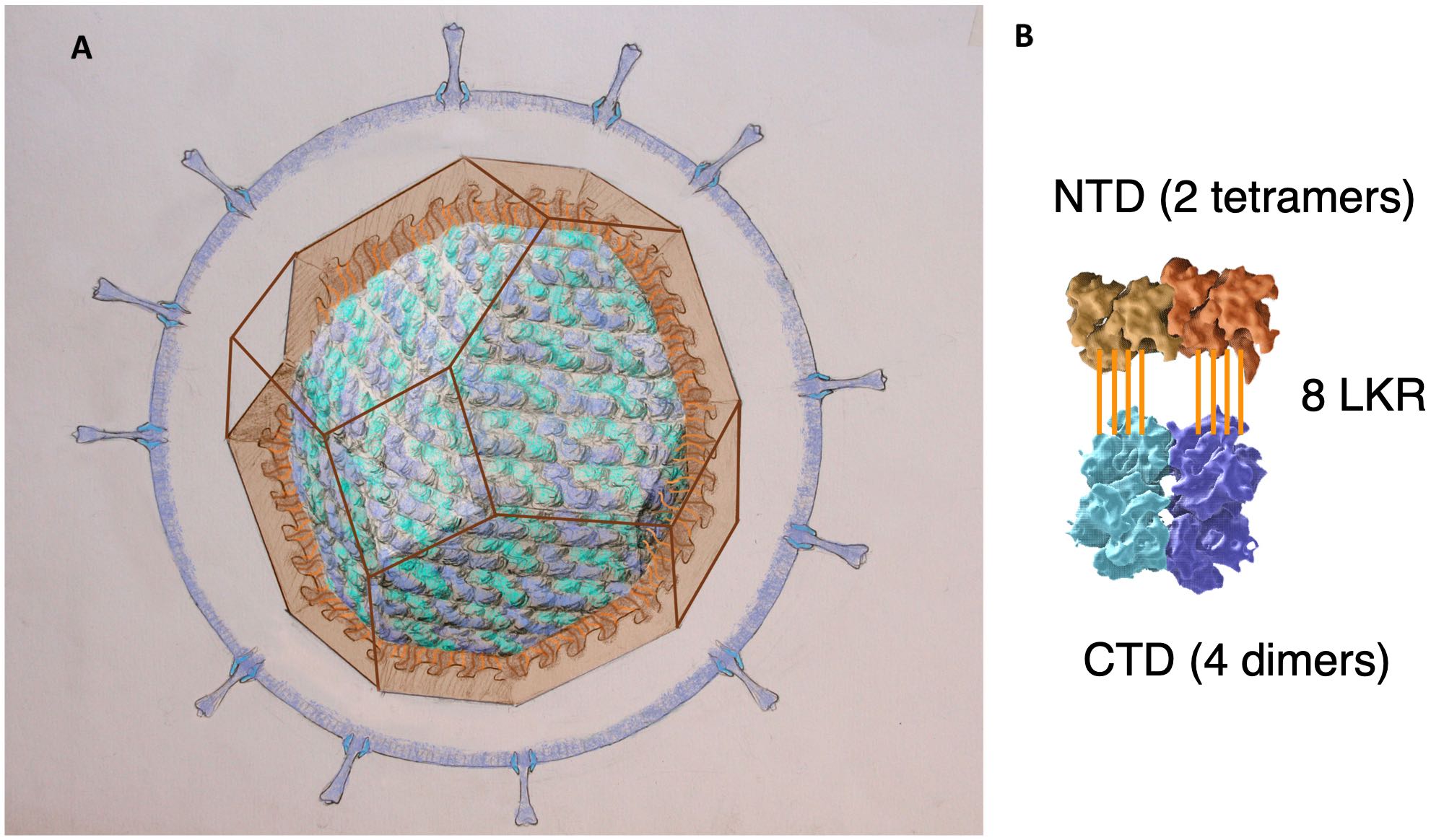}
	\caption{ Artistic rendering of the nucleocapsid structure. A) the membrane and the S and M proteins are depicted in blue colors. The external shell is represented with a transparent brown truncated octahedron. The interior shows the continuos coil packing of the blue and cyan CTD helices. B) High resolution structures of the NTD modules (2 tetramers) and CTD modules (4 dimers) linked by LKR regions.
	}
	\label{fig:crude}
\end{figure}
\medskip

Cryo electron tomography showed that the coronavirus nucleocapsid is separated from the envelope by a gap, which has revealed to contain thread-like densities that connect the protein density on the inner face of the viral membrane to a two-dimensionally ordered ribonucleoprotein layer \cite{Barcena:2009aa} .  Focal pairs revealed the existence of an extra internal layer that was attributed to the M protein, but that is compatible with the composite model we propose. Moreover, nucleoprotein densities were observed as a paracrystalline RNP shell, and may be partially organized at points of contact of the RNP lattice. The distribution of density in the viral core was consistent with a membrane-proximal RNP lattice formed by local approaches of the coiled ribonucleoprotein \cite{Barcena:2009aa, Neuman:2006aa} .  In the interior of the particles, coiled structures and tubular shapes were observed, consistent with a helical CTD model. The ribonucleoprotein appears to be extensively folded onto itself, assuming a compact structure that tends to closely follow the envelope at a distance of 4 nm \cite{Barcena:2009aa} . This indicates the existence of an additional layer that would confer the virion envelope its remarkable thickness.

\section*{Conclusions}

Viral nucleocapsids must obey a structural strategy that efficiently packs a long continuous chain of nucleic acid in an ordered manner, and thus it needs to adopt a consistent morphology. At the same time it is biologically required to have the versatility to disassemble and reassemble the components, in this case a tubular helicoidal packing of RNA and protein. We propose that these cannot be arbitrarily packed inside the virion but must be modulated, implying structural forms that are robustly built. The model we present is compatible with the known high resolution structures of the basic elements of N protein, their stoichiometry, and the lattice densities observed by cryoEM and cryo electron tomography. The octahedral geometry was previously observed in the bacteriophage MS2 (pdb 2VTU \cite{Plevka:2008aa}), a smaller nucleocapsid with no membrane. However, many EM studies did not observe a clear octahedral nucleocapsid as we propose, but rather describe roughly spherical pleomorphic particles. We propose that these observations can be reinterpreted under the current model. The apparent pleomorphism of CoVs may not be caused by RNP, but the transformations of the membrane that surrounds the nucleocapsid. If the capsid is octahedral, the membrane may flatten in the facets, giving the impression of strong deformations (Fig. 5). Recall that in CoVs the distance between the membrane and the nucleocapsid may reach 15nm of untidy regions, and thus this distance could vary between the vertices of the truncated octahedron and be much closer than the centers of the facets. These can be the places where N interacts with the M protein, a known necessary component for virion assembly. Coronavirus N proteins are appealing drug targets against coronavirus-induced diseases. A variety of compounds targeting the coronavirus nucleocapsid protein have been developed and many of these show potential antiviral activity \cite{Chang:2016aa} .

\medskip
	\begin{figure}
\centering
	\includegraphics[width=0.8\textwidth]{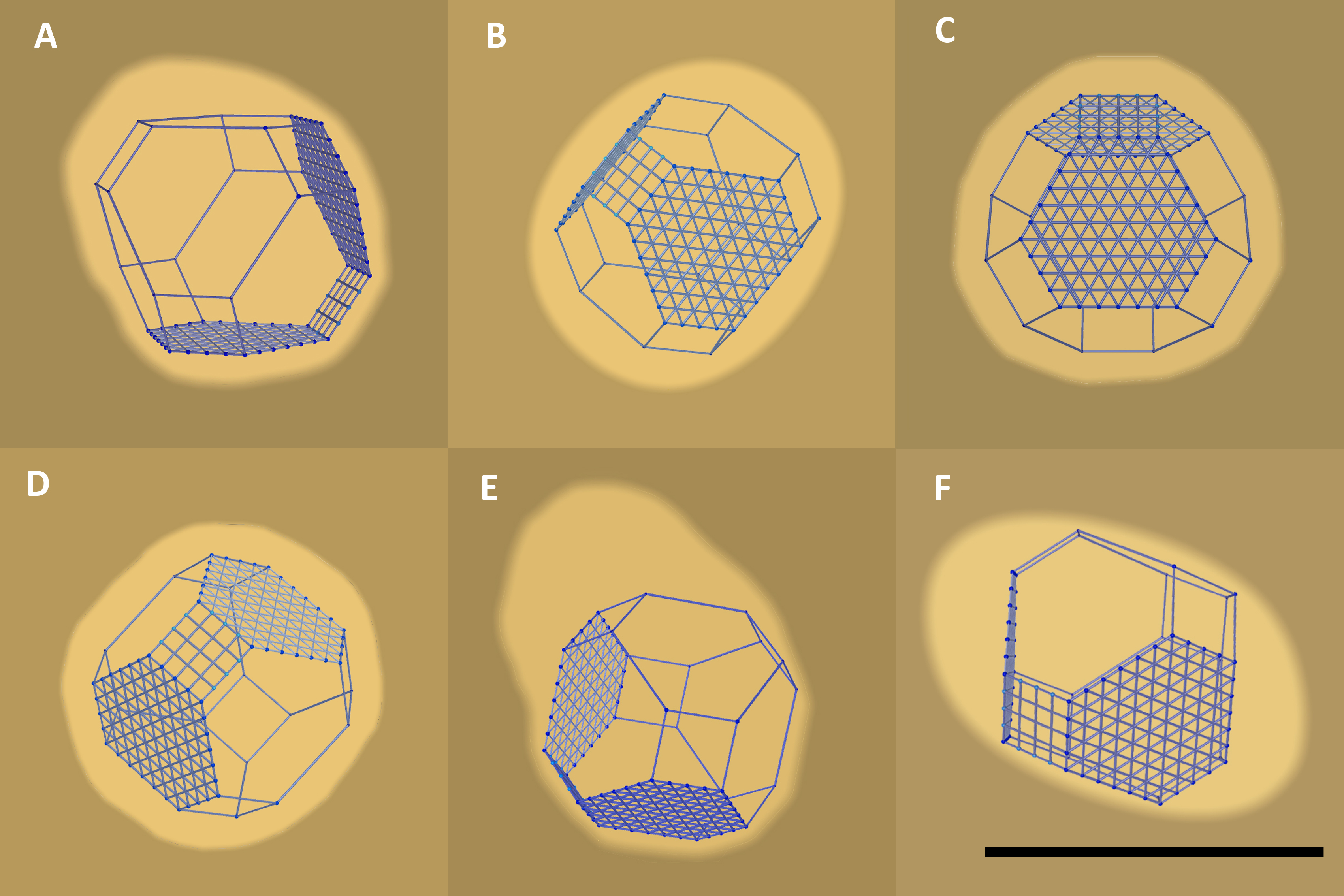}
	\caption{ Apparent pleomorphism of coronavirus. Different perspectives of the model are shown superimposed to the electron microscopy densities observed by Almeida et .al  \cite{combs2020}  in panels A) and E) and by Neuman et. al. \cite{Neuman:2006aa} in the other panels. Scale bar is 100nm.
	}
	\label{fig:crude}
\end{figure}
\medskip

\acknowledgement

This work was supported by the Consejo Nacional de Investigaciones Cient\'ificas y T\'ecnicas de Argentina (CONICET), the Agencia Nacional de Promoci\'on Cient\'ifica y Tecnol\'ogica (ANPCyT), the Universidad de Buenos Aires and NASA Astrobiology Institute. ADN and DUF are Career Investigators of CONICET.

%

%

%


\bibliography{Bibliography.bib}

\end{document}